# Superconducting nanowire single-photon detector with ultralow dark count rate using cold optical filters

## Hiroyuki Shibata[1,2], Kaoru Shimizu[1], Hiroki Takesue[1], and Yasuhiro Tokura[1]

[1]NTT Basic Research Laboratories, NTT Corporation, Atsugi-shi, Kanagawa 243-0198, Japan

[2]Nanophotonics Center, NTT Corporation, Atsugi-shi, Kanagawa 243-0198, Japan

We report the fabrication of a superconducting nanowire single-photon detector (SSPD or SNSPD) with an ultralow dark count rate. By introducing optical band-pass filters at the input of the SSPD and cooling the filters at 3 K, the dark count rate is reduced to less than 1/100 at low bias. An SSPD with 0.1 cps dark count rate and 5.6% system detection efficiency at 1550 nm wavelength is obtained. We show that a quantum key distribution (QKD) over 300 km of fiber is possible based on a numerical calculation assuming a differential phase shift QKD protocol implemented with our SSPDs.

Superconducting single-photon detectors (SSPDs or SNSPDs) based on ultrathin superconducting nanowires are now widely used as high-performance single-photon detectors in many fields such as quantum information and quantum optics.[1-3] SSPDs offer a wide detection wavelength range from ultraviolet to infrared, a high detection efficiency ($\eta$) of more than 10%, a low dark count rate (DCR) of 10 to 1000 cps, and a small timing jitter ($\Delta t$) of less than 100 ps. There have been many efforts to further improve their performance. In particular, many studies have been attempted to increase system $\eta$, and recently, a very high system $\eta$ of 93% has been reported using amorphous WSi.[4-6] Besides $\eta$, the DCR is also an important factor determining the performance of single-photon detectors. The figure of merit of a single-photon detector is usually represented as $\eta/(\mathrm{DCR}\Delta t)$.[7,8] In the figure of merit, decreasing DCR has the same weight as that of increasing $\eta$.

The importance of low DCR is also evident in long-distance quantum key distribution (QKD) experiments.[9] In QKD, the quantum bit error rate (QBER) increases as the length of the fiber increases. A secure key cannot be obtained from the sifted key when QBER exceeds a certain level, which limits the QKD distance. Since the QBER is usually governed by the DCR of the detector, a single-photon detector with low DCR is essential for long-distance QKD. Actually, to achieve the QKD distance of 260 km, the longest reported so far, SSPDs were operated in a very low-bias region to strongly reduce DCR (DCR = 1 cps, $\eta$ = 3%).[10] To further increase the distance of QKD, an SSPD with lower DCR is necessary.

In this letter, we describe the fabrication of an SSPD with a dark count rate below 1 cps.

The fabrication process of the device is as follows.[11,12] NbN thin films of 3.5 nm thickness were synthesized at 350°C on a MgO (100) substrate by reactive DC magnetron sputtering. The meander pattern of NbN was fabricated by the standard e-beam process using a negative resist and $C_2F_5$ dry etching. The size of the meander is 10 × 10 μm2 with a line and space width of 80 nm. Then, the cavity structure was formed on the meander using the standard photolithography process to enhance the detection efficiency. The cavity was composed of a 260-nm-thick $SiO_2$ layer, 2-nm-thick Ti, and 100-nm-thick Au. A device with $T_c$ = 10.5 K and $J_c$ = 5.3 × $10^6$ A/cm$^2$ at 4.2 K was obtained by the above process.

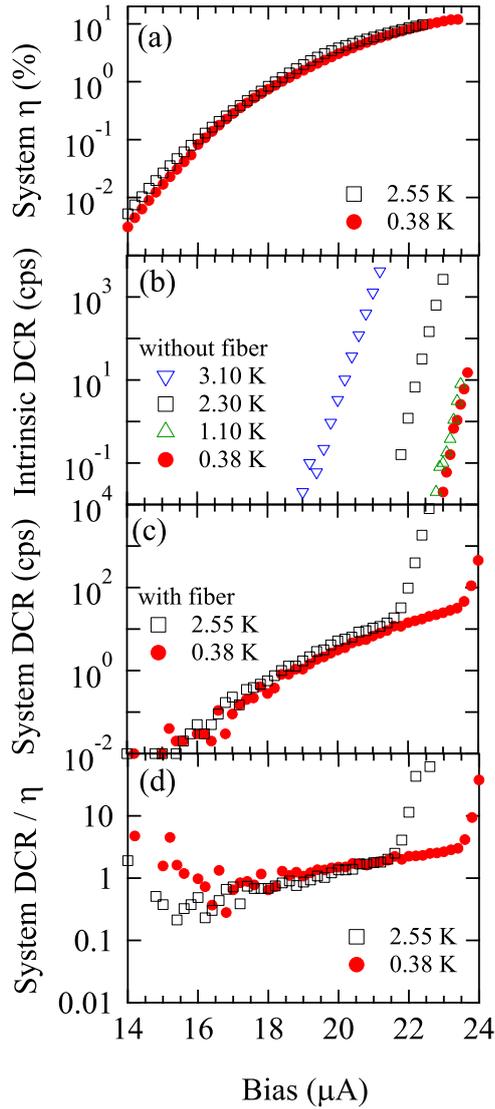

Fig. 1. Bias current dependences of (a) system detection efficiency ($\eta$), (b) intrinsic dark count rate (DCR), (c) system DCR, and (d) system DCR/$\eta$ of the device at various temperatures.

For the optical characterization of the device, we used a $^3$He cryocooler, which is able to cool the device down to 0.38 K. The stability of the temperature is better than 5 mK at the base temperature, while it increases to 50 mK at high temperature (0.5 – 3 K). An external 50 Ω shunt resistor was connected in parallel to the device on the cold stage and the signal was amplified using room-temperature rf amplifiers. A continuous-wave tunable laser with around 1550 nm wavelength was strongly attenuated and focused on the device to about 10 μm in diameter using a spherical lens at the end of the single-mode fiber.[13,14] Transmission spectra of the band-pass filters were measured down to $\lambda$ = 2.5 μm by using a long-wavelength optical spectrum analyzer (YOKOGAWA AQ6375) and an incoherent white light source (Anritsu MG922A).

Figure 1(a) shows the bias current dependence of the system $\eta$ of the device. At 0.38 K, the system $\eta$ increases as the bias current increases, and reaches 12% at the highest bias current of 23.4 μA. At 2.55 K, almost the same line is followed, except that the highest efficiency is reduced to 9.4% at 22.4 μA. Figure 1(b) shows the bias current dependence of the DCR without connecting the optical fiber to the device, indicating the intrinsic DCR of the device. Although the intrinsic DCR is large in the high-bias-current region, it exponentially decreases as the bias current decreases and becomes below 0.1 cps. The curve shifts to the high-bias region as the temperature decreases and saturates below 1 K. These features drastically change when an optical fiber is connected to the device without optical light input, as shown in Fig. 1(c). Although the exponential behavior is seen in the high-bias region, there appears a broad background in the low-bias region. As shown in Fig. 1(d), DCR/$\eta$ is almost constant in the low-bias region, suggesting that the origin of the broad background is not the intrinsic property of the detector. This broad background comes from blackbody radiation at room temperature, which propagates through the optical fiber. This indicates that the system DCR is governed by the blackbody DCR in the low-bias region. These features have also been reported in ref. 15.

The DCR due to a single-mode blackbody cavity at

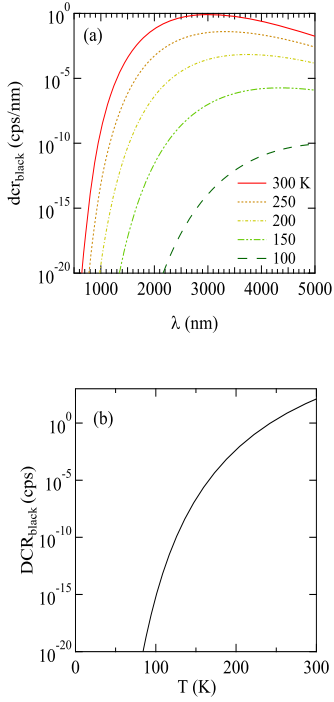

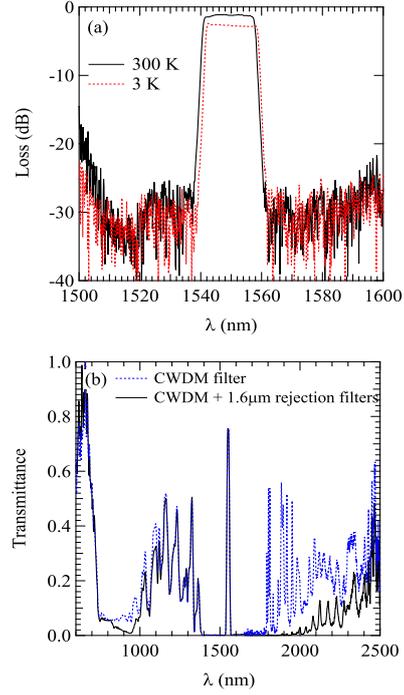

Fig. 2. (a) Calculated dark count rate ($dcr_{black}$) per nanometer wavelength due to blackbody radiation using eq. (1), and (b) $DCR_{black}(T)$ calculated by integrating $dcr_{black}$ between 200 to 2500 nm.

Fig. 3. (a) Transmission characteristics of the band-pass filters introduced at the input of the device at room temperature and at 3 K. (b) Transmission characteristics of band-pass filters (solid line) and only CWDM filter (dashed line) at room temperature in a wide wavelength region.

temperature T through single-mode optical fiber is approximately expressed as

$$DCR_{black}(\lambda_0, \lambda_1, T) = \int_{\lambda_1}^{\lambda_0} dcr_{black}(\lambda, T) d\lambda$$

$$= \int_{\lambda_1}^{\lambda_0} \frac{2c\varepsilon\eta}{\lambda^2} \frac{1}{e^{hc/\lambda k_B T} - 1} d\lambda, \quad (1)$$

where $dcr_{black}$ ($\lambda$,T) is the dark count rate per nanometer wavelength, ε is emissivity of the cavity, and $\lambda_0$ and $\lambda_1$ are the long- and short-wavelength absorption edges of the fiber.[16,17] Figure 2 shows wavelength dependence of $dcr_{black}$ ($\lambda$,T) at various temperatures, calculated using eq. (1). In the calculation, we assume that ε = 1 for all wavelengths, η = $10^2$exp(-2 × ln10 × $10^6$ × λ) below λ = 1000 nm, and η = 1 above λ = 1000 nm.[2] At 300 K, $dcr_{black}$ takes the maximum at around λ = 3000 nm and exponentially decreases at short wavelength. As the temperature decreases, $dcr_{black}$ strongly decreases and becomes negligible below T < 100 K. The inset of Fig. 2 shows the calculated temperature dependences of $DCR_{black}$. Here, we take $\lambda_1$ = 200 nm and $\lambda_2$ = 2500 nm as the transmission range of the 1.5 μm single-mode fiber. At 300 K, $DCR_{black}$ becomes 1.28 × $10^2$ cps, which is comparable to the observed DCR in Fig. 1(c), but it exponentially decreases as temperature decreases. This indicates that DCR from the blackbody can be removed by introducing an optical band-pass filter and cooling it below 100 K.

For the cold band-pass filters, a coarse wavelength division multiplexer (CWDM) filter and 1.6 μm rejection short-wavelength pass filter, which are commercially available, are used.[18] They are connected in series and are cooled at the second stage of the cryocooler cooled to 3 K.

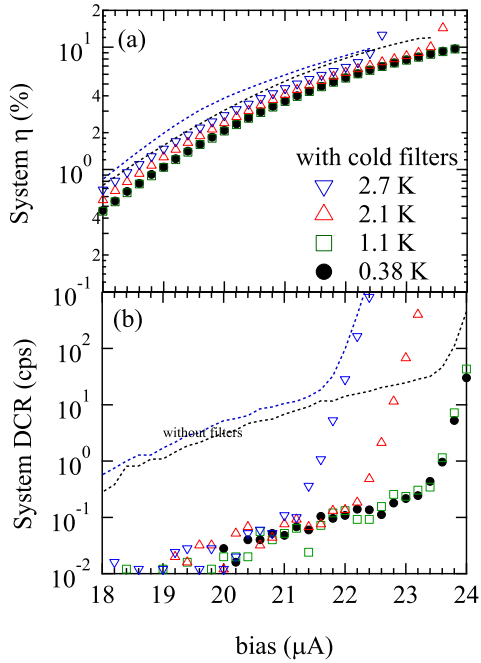

Fig. 4. Bias current dependence of (a) system detection efficiency (η) and (b) system dark count rate (DCR) with filter that was cooled to 3 K at various device temperatures. The thin dashed lines are without cold filters at device temperatures of 0.38 (black) and 2.55 K (blue).

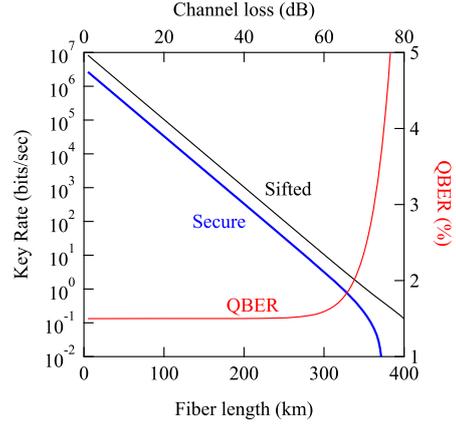

Fig. 5. Calculated fiber length dependences of key rate and QBER using the ultralow dark count SSPD (η = 5%, DCR = 0.1 cps, 1 GHz clock speed, 100 ps time windows, DPS-QKD protocol, 0.2 dB/km fiber loss).

Figure 3(a) shows the transmission properties of the filters. They have a bandwidth of 20 nm at 1550 nm wavelength and insertion loss of 1.2 dBm at room temperature, including the coupling loss at the cryocooler. As the temperature decreases to 3 K, the insertion loss increases to 2.7 dBm and the pass-band shifts about 1 nm. Broadband transmission properties of the filters are shown in Fig. 3(b). The long-wave blocking limit is about 1800 nm for a single CWDM filter, and it is extended to 2000 nm by adding a 1.6 μm rejection filter.

Figure 4 shows (a) system η and (b) system DCR of the device with cold filters at 3 K. The system η decreases to 60% due to the loss of the filters. Although the system DCR strongly increases in the high-bias region, it strongly decreases in the low-bias region due to the protection of blackbody radiation. At a bias current of 22.0 μA and at 0.38 K, system η decreases from 8.1 to 5.6%, while system DCR decreases from 14 to 0.1 cps by the insertion of cold filters, and the figure of merit increases 97-fold. This clearly shows the importance of cold filters to the figure of merit. It is noted that such a high increase of the figure of merit cannot be expected by increasing η of the device without insertion of cold filters, since $DCR_{black}$ also increases with increasing η. The strong decrease of system DCR is also observed at 2.7 K, which can be reached using the standard Gifford-McMahon cryocooler. At a bias current of 21.0 μA and at 2.7 K, system η of 4.6% and system DCR of 0.1 cps are obtained by the insertion of cold filters.

Although the DCR is strongly reduced by the cold filters, there remains a broad background on the order of 0.1 cps in the low-bias region in Fig. 4(b). This is due to the remaining blackbody radiation that passes through the cold filters. As shown in Fig. 3(b), there are three main pass-bands: shorter than 1400 nm, 1540 to 1560 nm, and longer than 2000 nm. Since $dcr_{black}$ increases at the longer wavelength, the main source of residual DCR comes from the pass-band longer than 2000 nm. Actually, we confirmed that system DCR decreases to only 1/5 with the CWDM filter alone. It seems possible to further reduce the $DCR_{black}$ by using a cold filter with a wider rejection-wavelength region.

Figure 5 shows the calculation of the fiber length dependence of key rate and QBER using the above SSPD. In the calculation, we assume a standard telecom fiber with the loss coefficient of 0.2 dB/km, the differential phase shift QKD (DPS-QKD) protocol under general individual attacks, a 1 GHz clock rate, 100 ps time windows, $\eta = 5\%$, and DCR = 0.1 cps.[19,20] In the figure, the sifted key rate exponentially decreases as the fiber length increases. The secure key rate also exponentially decreases but it drops at about 350 km due to the strong increase of QBER. The secure key rate is 2.5 bit/s at 300 km, which is enough for secure communications.

In conclusion, we report the fabrication of SSPD with an ultralow dark count rate. The main origin of the dark count rate in the low-bias region is blackbody radiation at room temperature, which propagates through optical fiber. By introducing cold band-pass filters at the input of the SSPD, the dark count rate can be strongly reduced. An SSPD with a 0.1 cps dark count rate and 5.6% system detection efficiency is obtained, which can be used in QKD over a distance of more than 300 km.